\documentclass[12pt]{article}
\usepackage{putex}
\usepackage{graphicx}
\begin{document}

\preprint{PUPT-2321}

\institution{PU}{Joseph Henry Laboratories, Princeton University, Princeton, NJ 08544, USA}

\title{Peculiar properties of a charged dilatonic black hole in $AdS_5$}

\authors{Steven S. Gubser and Fabio D. Rocha}

\abstract{We study a charged dilatonic black hole in $AdS_5$, derived from a lagrangian involving a gauge field whose kinetic term is modified by the exponential of a neutral scalar.  This black hole has two properties which one might reasonably demand of the dual of a Fermi liquid: Its entropy is proportional to temperature at low temperature, and its extremal limit supports normal modes of massless, charged bulk fermions.  The black hole we study has a simple analytic form because it can be embedded in type~IIB string theory as the near-horizon limit of D3-branes with equal spins in two of the three independent transverse planes.  Two further properties can be deduced from this embedding: There is a thermodynamic instability, reminiscent of ferromagnetism, at low temperatures; and there is an $AdS_3$ factor in the extremal near-horizon geometry which accounts for the linear dependence of entropy on temperature.  Altogether, it is plausible that the dilatonic black hole we study, or a relative of it with similar behavior in the infrared, is the dual of a Fermi liquid; however, the particular embedding in string theory that we consider is unlikely to have such a dual description, unless through some unexpected boson-fermion equivalence at large $N$.}

\date{November 2009}

\maketitle

\tableofcontents

\section{Introduction}
\label{INTRODUCTION}

The AdS/CFT correspondence allows us to better understand some aspects of strongly coupled field theories by studying simple gravitational models and significant work has gone into applying this approach to the study of strongly coupled systems at finite temperature and density (see \cite{Herzog:2007ij}  for early work). The $AdS_4$-Reissner-Nordstrom black hole was one of the first gravity models to be considered in this context \cite{Hartnoll:2007ih,Hartnoll:2007ip,Hartnoll:2007ai}. This black hole background is simple because it involves only the metric and a gauge field, and it is reliable in the supergravity approximation because it has curvatures that can be made everywhere small.  In fact, the extremal $AdS_4$-Reissner-Nordstrom solution (hereafter RNAdS) is a charged domain wall interpolating between $AdS_4$ in the ultraviolet and $AdS_2 \times {\bf R}^2$ near the horizon.  The domain wall structure has been particularly important in the study \cite{Liu:2009dm,Cubrovic:2009ye,Faulkner:2009wj} of two-point functions of an operator ${\cal O}_\psi$ dual to a charged fermion $\psi$ in the bulk.  These calculations (all of them relying to some degree on numerics) point to the existence of an isolated fermion normal mode at finite wave-number, call it $k_F$.  The authors of \cite{Liu:2009dm,Cubrovic:2009ye,Faulkner:2009wj} have argued that this normal mode signals a Fermi surface.  The argument runs roughly as follows.  RNAdS carries a charge under a $U(1)$ symmetry which is gauged in the bulk and global on the boundary.  Fermions in the boundary theory can plausibly be assumed to be charged under this $U(1)$.  The zero-temperature state at finite charge density which is dual to RNAdS is supposed to be a Fermi liquid.  ${\cal O}_\psi$ is assumed to have some overlap with the operator which creates one fermion in the boundary field theory.  If this fermion is created very near the Fermi surface, then it should have a long lifetime.  So the spectral measure should have a spike at energy equal to the Fermi energy and momentum equal to the Fermi momentum.  Now, the Fermi energy can reasonably be identified as the charge of the bulk fermion times the potential difference between the boundary of $AdS_4$ and the black hole horizon, because this is the amount of energy it takes to add one bulk fermion's worth of charge to the black hole.\footnote{Implicit in the argument at this step is that the boundary field theory fermion has the same $U(1)$ charge as the bulk fermion.  This would be true, for example, if ${\cal O}_\psi = \tr X \lambda$ where $X$ is a neutral scalar and $\lambda$ is the boundary field theory fermion.} In the conventions of \cite{Liu:2009dm,Faulkner:2009wj}, which we also adopt, energy equal to the Fermi energy corresponds to frequency equal to $0$.  The spike found in the spectral measure through explicit calculations on the gravity side is indeed at zero frequency.  Furthermore, it had previously been pointed out in \cite{Rey:2008zz,Sin:2009rf} that some of the thermodynamical and transport properties of near extremal RNAdS are also suggestive of a Fermi surface. So everything makes sense.

Or does it?  The elephant in the room is the macroscopic zero-temperature entropy of RNAdS, which seems at odds with a description of the zero-temperature dual as a degenerate Fermi liquid.  For the sake of a simple discussion, let's focus on the $AdS_5$ case.  The dual is then a gauge theory (at least in all known constructions) for which the ranks of the gauge groups are large.  Let the rank of one gauge group be $N$.  Then the total entropy density is $N^2 k_F^3$ up to factors of order unity.  More conventionally, one could express entropy density as $N^2 \Omega^3$ times factors of order unity, where $\Omega$ is the chemical potential for the $U(1)$ symmetry; but $k_F / \Omega$ is $O(1)$.  One way out is to suppose that the zero-point entropy owes to some unspecified dynamics of the colored degrees of freedom, while the Fermi liquid dynamics is a subleading effect in $N$.  This seems in line with the calculations of one-loop bulk effects in \cite{Denef:2009yy}.  The picture that seems to emerge is that the number of fermions in the Fermi liquid (or at least, the number of fermions near the edge of the Fermi sea) is $O(1)$ rather than $O(N^2)$.  Perhaps they should be thought of as color singlet bound states of an adjoint scalar and an adjoint fermion, created by an operator of the form $\tr X\lambda$, where $X$ is the scalar and $\lambda$ is the fermion.  It is remarkable that there should be such a bound state when the entropy indicates that the theory as a whole is in a deconfined state.  What seems really odd, if one subscribes to this picture, is that the spectral measure of the two-point function of ${\cal O}_\psi$ scales naturally as $N^2$.  To see this without committing oneself to a definite form of ${\cal O}_\psi$, consider that the two-point function of the stress-tensor of the boundary field theory certainly scales as $N^2$, as does the two-point function of the operator dual to the dilaton: indeed, this scaling, and the agreement of the overall coefficient in certain cases, provided early hints of AdS/CFT \cite{Klebanov:1997kc,Gubser:1997yh,Gubser:1997se}.  Likewise the supercurrent, dual to the bulk gravitino, has two-point functions that scale as $N^2$, and so do all single-trace, color-singlet operators dual to bulk supergravity fields, simply because the on-shell action that defines the two-point functions naturally includes a prefactor of $1/G_5$, which is proportional to $N^2$.  It is hard to see how this scaling squares with the picture of the spike in the spectral measure owing to quasi-particle dynamics of a color-singlet bound state: the magnitude of the spike itself scales as $N^2$.  In summary, the picture of $O(1)$ Fermi liquid phenomena resting on top of deconfined $O(N^2)$ dynamics of colored degrees of freedom presents some serious puzzles, and pending a resolution of them---including a microscopic account of the zero-point entropy with coefficients that agree or almost agree between field theory and gravity---one should feel entitled to some doubt about the whole picture.

Matters would be simpler if the zero-point entropy weren't there.  Better yet would be if the $O(N^2)$ thermodynamics also exhibited linear specific heat, as one expects for a Fermi liquid.  If such a setup could be found, with a normal mode similar to the one that exists for RNAdS, then one could plausibly advance the interpretation that adjoint fermions in the field theory are in a Fermi liquid state at zero temperature; that the normal mode (with magnitude scaling as $N^2$) signals the existence of $O(N^2)$ quasi-particle excitations (adjoint or bifundamentally charged fermions in quiver theories) at the edge of the Fermi surface; and that the specific heat at low temperatures (also scaling as $N^2$) is the specific heat of the Fermi liquid.

So, can a suitable black hole be constructed in an asymptotically $AdS_5$ geometry?  The answer is ``Yes.''  In fact one can even concoct a theory where the black hole solution is analytically known.  The simplest such theory (at least, the simplest one we know) is
 \eqn{Lred}{
  {\cal L} = {1 \over 2\kappa^2} \left[ R - {1 \over 4} e^{4\alpha} F_{\mu\nu}^2 - 
    12 (\partial_\mu\alpha)^2 + {1 \over L^2} (8e^{2\alpha} + 
      4e^{-4\alpha}) \right] \,,
 }
and the spatially uniform, electrically charged solution is
 \eqn{TwoChargeExpress}{
  ds^2 &= e^{2A} (-h \, dt^2 + d\vec{x}^2) + {e^{2B} \over h} dr^2 \qquad
    A_\mu dx^\mu = \Phi dt  \cr
  A &= \log {r \over L} + {1 \over 3} \log\left( 1 + {Q^2 \over r^2} \right) \qquad
  B = -\log {r \over L} - {2 \over 3} \log\left( 1 + {Q^2 \over r^2} \right)  \cr
  h &= 1 - {\mu L^2 \over (r^2 + Q^2)^2} \qquad
  \Phi = {Q\sqrt{2\mu} \over r^2 + Q^2} - {Q\sqrt{2\mu} \over r_H^2 + Q^2}  \cr
  \alpha &= {1 \over 6} \log\left( 1 + {Q^2 \over r^2} \right) \,.
 }
This black hole is extremal if $r_H = 0$, which implies $\mu L^2 = Q^4$.  The extremal solution has a naked singularity at $r=0$, which we will say more about in section~\ref{GIANT}.  The neutral scalar $\alpha$ plays the role of a dilaton because it controls the physical gauge coupling.

The thermodynamics of the charged dilatonic black hole \eno{TwoChargeExpress} is most easily expressed in the microcanonical ensemble in terms of a rescaled energy density, entropy density, and charge density:
 \eqn{RescaledDensities}{
  \hat\epsilon \equiv {\kappa^2 \over 4\pi^2 L^3} \epsilon = 
    {3\mu \over 8\pi^2 L^6} \qquad
  \hat{s} \equiv {\kappa^2 \over 4\pi^2 L^3} s = {r_H \sqrt\mu \over 2\pi L^5} \qquad
  \hat\rho \equiv {\kappa^2 \over 4\pi^2 L^3} \rho = 
    {Q \sqrt{2\mu} \over 4\pi^2 L^5} \,.
 }
From \eno{RescaledDensities} together with the condition $h(r_H) = 0$, one may straightforwardly verify the micro-canonical equation of state:
 \eqn{eos}{
  \hat\epsilon = {3 \over 2^{5/3} \pi^{2/3}} \left( \hat{s}^2 + 
    2\pi^2 \hat\rho^2 \right)^{2/3} \,.
 }
The temperature and chemical potential can be found by differentiation:
 \eqn{OmegaT}{
  T = \left( {\partial\hat\epsilon \over \partial\hat{s}} \right)_{\hat\rho} \qquad
  \Omega = \left( {\partial\hat\epsilon \over \partial\hat\rho} \right)_{\hat{s}} \,.
 }
One easily finds
 \eqn{LinearS}{
  \hat{s} = \pi \sqrt{2\hat\epsilon \over 3} \, T \approx (\pi^2 \hat\rho)^{2/3} \, T
    \approx {\Omega^2 \over 4} T \,,
 }
where the approximate equalities hold in the low-temperature limit.  The rescaled specific heats at constant charge density and constant chemical potential,
 \eqn{SpecificHeat}{
  \hat{C}_{\hat\rho} = T \left( {\partial\hat{s} \over \partial T} \right)_{\hat\rho}
    \qquad
  \hat{C}_\Omega = T \left( {\partial\hat{s} \over \partial T} \right)_\Omega
 }
coincide with each other and with the rescaled entropy density $\hat{s}$ in this limit.

With a black hole in hand whose low-temperature thermodynamics lends itself to the claim that the dual is a Fermi liquid, the next obvious question is whether it supports isolated fermionic normal modes similar to the ones found in RNAdS.  We will show by example in section~\ref{FERMION} that it does.  Next one might inquire whether one can embed this black hole in string theory.  Indeed one can.  In fact, \eno{Lred} is a consistent truncation of maximal gauged supergravity in five dimensions \cite{Gunaydin:1985cu}, and the solution \eno{TwoChargeExpress} is the black hole solution in this theory where two of the three commuting $U(1)$ gauge groups carry equal charge, and the third carries no charge.  Thus it can be immediately lifted to a ten-dimensional geometry, asymptotic to $AdS_5 \times S^5$, which describes the near-horizon limit of D3-branes which have equal spin in two of the three orthogonal transverse planes, and zero spin in the third plane \cite{Cvetic:1999xp}.  By way of comparison, RN$AdS_5$ can be lifted to the near-horizon limit of D3-branes with equal spin in all three orthogonal transverse planes.

When \eno{TwoChargeExpress} is lifted to an asymptotically $AdS_5 \times S^5$ solution to type~IIB supergravity, its field theory dual must be ${\cal N}=4$ super-Yang-Mills theory (hereafter ${\cal N}=4$ SYM), which has massless scalars charged under the $U(1)$ symmetry dual to $A_\mu$, as well as massless charged fermions.  It is hard to see how these charged scalars would fail to take over the dynamics at low temperatures and finite chemical potential: in particular, one would expect that they condense, spontaneously breaking the $U(1)$ symmetry that is gauged in the bulk theory \eno{Lred}.  This indeed happens, as can be seen from studying the dynamics of bulk scalars in the $20$ of $SO(6)$ in the background \eno{TwoChargeExpress} \cite{TwentyUnpublished}.  In section~\ref{UNSTABLE} we note another, simpler instability: There is a negative thermodynamic susceptibility below a critical temperature which leads to the spontaneous breaking of an $SU(2)$ symmetry as well as translational invariance.  This negative susceptibility is a well studied \cite{Gubser:1998jb,Cvetic:1999rb} example of the Gregory-Laflamme instability \cite{Gregory:1993vy,Gregory:1994bj}, although its interpretation here in terms of breaking an $SU(2)$ symmetry is new as far as we know.

While the lift to an asymptotically $AdS_5 \times S^5$ geometry does not particularly encourage the view that the dual physics is a Fermi liquid, it does provide an interesting explanation of the linear specific heat: there is an $AdS_3$ factor in the near horizon geometry, which takes two of its dimensions from the $AdS_5$ factor and one from the $S^5$.  In section~\ref{GIANT}, we explain the appearance of this $AdS_3$ factor heuristically in terms of an effective string built from intersecting giant gravitons, along the lines of \cite{Gubser:2004xx}.  It is tempting to speculate that the non-chiral conformal invariance associated with the $AdS_3$ geometry is a more general feature of embeddings of black holes with linear specific heat in string theory and M-theory.  The question naturally arises: Could there be an embedding where there is a Fermi liquid in the dual field theory which somehow explicitly realizes the non-chiral conformal symmetry?

The spinning D3-branes construction has a well-known M-theory analogue. In section~\ref{FOUR}, we find that it can reproduce many of the features we found appealing in the string theory construction. Namely, it has a four-dimensional reduction consisting of a charged dilatonic black hole that supports an isolated fermion normal mode at finite $k$ at zero temperature and has linear specific heat at low temperatures.

Once one has achieved a detailed understanding of a charged dilatonic black hole with linear specific heat at low temperature, a natural follow-up question is whether the theory \eno{Lred} can be modified in a simple way to accommodate other behaviors for the specific heat.  We explore this question in section~\ref{SCALING}, exhibiting some further exactly solvable examples and summarizing their thermodynamic properties.

We end with a summary of our findings in section~\ref{SUMMARY}.

\section{Fermion normal modes}
\label{FERMION}

Let us now consider fermions in the zero temperature limit of the charged dilatonic black hole \eno{TwoChargeExpress}. We will take the fermionic action to be
 \eqn{FermionicS}{
  S_f = i\int d^5 x\,\bar\psi (\Gamma^\mu D_\mu - m) \psi +S_{\rm bdy}\,
 }
where $D_\mu \psi = \left( \partial_\mu + {1 \over 4} \omega_\mu{}^{\underline{\rho\sigma}} \Gamma_{\underline{\rho\sigma}} - i q A_\mu \right) \psi$ and $\omega_\mu{}^{\underline{\rho\sigma}}$ is the spin connection.  $S_{\rm bdy}$ is a boundary term necessary to have well defined variational problem \cite{Iqbal:2009fd} and does not affect the equations of motion. We will specify it below. We denote curved space indices by $\mu$ and tangent space indices by $\underline\mu$. It is convenient to chose a basis for the  $\Gamma$ matrices where $\Gamma^{\underline r}$ is diagonal, and we choose
\eqn[c]{Gammadef}{
\Gamma^{\underline{0}} = i\begin{pmatrix} \sigma_2 &0 \\ 0& \sigma_2\end{pmatrix} \,,\quad
\Gamma^{\underline{1}} = \begin{pmatrix} 0 &\sigma_1 \\  \sigma_1 &0\end{pmatrix} \,,\quad
\Gamma^{\underline{2}} = \begin{pmatrix} 0 &\sigma_3 \\ \sigma_3& 0 \end{pmatrix} \,,\cr
\Gamma^{\underline{3}} = i\begin{pmatrix} 0 &-1 \\ 1& 0\end{pmatrix} \,,\quad
\Gamma^{\underline{r}} = \begin{pmatrix} 1 &0 \\ 0& -1\end{pmatrix} \,,
}
where $\sigma_1$, $\sigma_2$, and $\sigma_3$ are the Pauli matrices.  The matrices $\Gamma^{\underline 0},\ldots,\Gamma^{\underline{3}}$ are then four-dimensional $\Gamma$ matrices with $\Gamma^5=\Gamma^{\underline{r}}$.

The action \eno{FermionicS} is {\it ad hoc} in the sense that we do not derive it from supergravity or from an embedding of the black hole solution \eno{TwoChargeExpress} in string theory.  We will explain such an embedding in sections~\ref{UNSTABLE} and~\ref{GIANT}, and in principle we could replace $S_f$ by the quadratic fermion action of maximally supersymmetric gauged supergravity in five dimensions.  However, this action is complicated, involving mixing between spin-$3/2$ fields and spin-$1/2$ field through the super-Higgs mechanism, Yukawa couplings, and also couplings between the gauge field strength $F_{\mu\nu}$ and bulk fermion bilinears.  So it seems to be quite a challenge to diagonalize the action.  Also, we do not want to commit ourselves to the embedding of \eno{Lred} into maximal gauged supergravity as the only potentially interesting one.  Altogether it seems worthwhile to start with the fermionic action \eno{FermionicS}.

We will only consider $m\geq 0$. Noting the translation symmetry in the $t,x^i$ directions, we take the fermion wave-function to be of the form
\eqn{PsiAnsatz}{
\psi(t,x^i,r) = e^{-i\omega t + i k x^1} u(r)= e^{-i\omega t + i k x^1} 
  \begin{pmatrix} u^+_1(r) \\ u^+_2(r) \\ u^-_1(r) \\ u^-_2(r) \end{pmatrix} \,.
}
Because of rotational invariance in the $\vec{x}$ direction, we can assume that the momentum is in the $+x^1$ direction: that is, $k>0$.  With this ansatz, Dirac's equation can be written as 
\eqn{Diracb}{
\left[ e^{-B}\sqrt{h}\Gamma^{\underline{r}} \partial_r +  ie^{-A}\left(k \Gamma^{\underline{1}} - \frac{\omega +q \Phi}{\sqrt h} \Gamma^{\underline{0}}\right) +\frac{8 h A' +h'}{4\sqrt h e^B}\Gamma^{\underline{r}} 
- m\right] u =0 \,.
}
To find fermion normal modes we need to solve \eqref{Diracb} with the appropriate boundary conditions at $r=0$ and look for values of the parameters for which $u$ grows more slowly as $r\to+\infty$ than the generic solution.

To decide on the boundary conditions at $r=0$, we can solve \eqref{Diracb} near $r=0$ using series expansions. The form of the expansion depends crucially on whether $\omega$ is zero or non-zero. We expect normal modes only at $\omega=0$, since for $\omega \neq0$ the solutions exhibit a non-zero flux in the $r$ direction, so let us consider the zero frequency case. The series expansion are then of the form
\eqn{HorSeriesZF}{
u = U r^{-{7\over 6} + {|k|\over|\Omega|}}\bigg(1+O\big(r^{1/3}\big)\bigg)+V r^{-{7\over 6} - {|k|\over |\Omega|}}\bigg(1+O\big(r^{1/3}\big)\bigg) \qquad (\omega=0)\,,
}
where $\Omega=\sqrt{2} Q/L^2$ is the chemical potential at zero temperature. The subleading terms have series expansions in integer powers of $r^{1/3}$ that can easily be found. The equation of motion also allows us to write the $3,4$ components of the constant spinors $U$ and $V$ algebraically in terms of their $1,2$ components, but the exact form of that relationship won't be important. The boundary condition that we will impose is to set $V$, the coefficient of the most divergent solution, to zero. This amounts to requiring that $\sqrt{-g} \bar\psi\psi$ be finite at $r=0$, a property shared by the purely infalling solutions for $\omega\neq 0$. We found numerically that purely infalling solutions for $\omega\neq 0$ approach the zero frequency solutions with $V=0$  as you take $\omega\to 0$.

Let us now consider the behavior of the solutions near the boundary, i.e., as $r\to +\infty$. In this limit, the geometry is $AdS_5$ with radius of curvature $L$ and the gauge field $\Phi$ is a constant, so we can solve \eqref{Diracb} , obtaining
\eqn{uInAdS}{
u^{\pm}_a = C^{\pm}_a e^{-{5\over2} A}  X_{\mp{1\over 2} -m L}\Big(\kappa e^{-A}\Big) + D^{\pm}_a e^{-{5\over2} A}  X_{\pm{1\over 2} +m L}\Big(\kappa e^{-A}\Big) 
}
where
\eqn{KappaDef}{
\kappa^2 \equiv L\sqrt{k^2- (q\Phi+\omega)^2} \,,
}
and $X_\nu$ is an appropriate Bessel function. If $\kappa^2>0$, we should take for $\kappa$ the positive square root and set $X_\nu=I_\nu$, the modified Bessel function. If $\kappa^2<0$, we can either take $\kappa=i\sqrt{(q\Phi+\omega)^2-k^2}$ or, equivalently, replace $\kappa$ by $|\kappa|$ in \eno{uInAdS} and set $X_\nu = J_\nu$.\footnote{We are assuming that $\nu \notin\mathbb{Z}$, so that $J_\nu$ and $J_{-\nu}$ are linearly independent. For $\nu\in\mathbb{Z}$, the $X_\nu$ should be built out of $J_\nu$ and $Y_\nu$. } Similarly to what happened for $r=0$, we can use the equation motion to write $C^-_a$ and $D^-_a$ algebraically in terms of $C^+_a$ and $D^+_a$, and are left with 4 independent constants of integration.

Expanding \eqref{uInAdS} at large $r$, we see that $u^+_a \propto C^+_a r^{-2+m L}$, and  we therefore identify normal modes with solutions of \eqref{Diracb} with $V=0$ for which $C^+_1=C^+_2=0$, i.e., for which the coefficient of the most divergent solution near the boundary is zero. Note that the structure of \eqref{Diracb} means that $u^+_1$ only couples to $u^-_2$ and $u^+_2$ only to $u^-_1$, so it is consistent to set $u^+_2=u^-_1=0$ (or $u^+_1=u^-_2=0$) and look simply for zeros of $C^+_1$ (or, resp.\ $C^+_2$). These normal modes correspond to poles of the retarded Green's function \cite{Iqbal:2009fd}, with normal modes with nonzero  $u^+_1$ (or nonzero $u^+_2$) corresponding to poles of $G_{11}$  (or poles of $G_{22}$, resp.). More concretely, if we choose the boundary term in the action \eqref{FermionicS} to be
\eqn{Sbdydef}{
S_{\rm bdy} = -i \int\limits_{r=1/\varepsilon} d^4 x\, \sqrt{-g g^{rr}} \bar\psi_+ \psi_-\,, \qquad \psi_\pm\equiv {1\over 2} \left(1\pm\Gamma^{\underline r}\right)\psi \,,
}
where $\varepsilon$ is a positive quantity to be taken to zero after functional derivatives are taken, then $\psi$ is dual to a positive chirality Weyl spinor\footnote{We could have made a different choice for $S_{\rm bdy}$, with $\psi_+ \leftrightarrow \psi_-$. With this choice, $\psi$ is dual to a negative chirality Weyl spinor, and the formula for the Green's function is slightly different.} and the retarded Green's function is given by
\eqn{GotGR}{
G_R = \left(\kappa \over 2\right)^{2mL} {\Gamma\left({1\over2} - m L\right)\over\Gamma\left({1\over2} + m L\right)} \begin{pmatrix} - i \frac{D^-_2}{C^+_1} &0 \\ 0 & i \frac{D^-_1}{C^+_2} \end{pmatrix} \,.
}

Before giving a concrete example of a normal mode, let's consider two heuristic constraints on suitable values of $k$.  First, we expect that $\kappa^2<0$, which for $\omega=0$ reduces to 
\eqn{kLimBdy}{
|k| < k_\textrm{max} \equiv L |q \Omega| \,.
}
We arrive at this condition by noting that if $\kappa^2>0$, then for a normal mode at large $r$, $u^+_a \propto e^{-{5\over 2} A} I_{{1\over 2} +m L}\Big(\kappa e^{-A}\Big)$, and so $u^+_a$ increases monotonically (and eventually exponentially in $e^{-A}$) as one goes down into $AdS_5$.  It is hard to see how this asymptotic behavior would match onto a solution that is normalizable at $r=0$.  On the other hand, if $\kappa^2<0$, $u^+_a \propto e^{-{5\over 2} A} J_{{1\over 2} +m L}\Big(|\kappa| e^{-A}\Big)$ at large $r$, which is oscillatory and there is no apparent obstruction to the matching.
Because the asymptotic behaviors we described are only approximations to $u$, one cannot take the bound \eno{kLimBdy} as rigorous.  

Second, we prefer to investigate wave-numbers $k$ such that the leading power in \eqref{HorSeriesZF} is positive, i.e.\ obeying
\eqn{kLimHor}{
|k|>k_\textrm{min} \equiv  {7 \over 6} |\Omega| \,. 
}
This condition simply ensures that $\psi$ does not diverge at $r=0$.  While this is not strictly necessary, it is convenient numerically and helps ensure the absence of back-reaction.  Conditions \eqref{kLimHor} and \eqref{kLimBdy} have nonzero intersection only if $q>q_\textrm{min}\equiv {7 \over 6L}$.  So, as a specific example, we consider the following choice of parameters:
 \eqn{ParameterChoice}{
  L=1 \qquad m = 0 \qquad Q = 1 \qquad q=2 \,.
 }
$L=1$ is a choice of units.  $Q=1$ corresponds to $\Omega=\sqrt{2}$, and $q=2$ satisfies the bound $q > q_\textrm{min}$.

\begin{figure}[h!]
  \centering\includegraphics[width=5in]{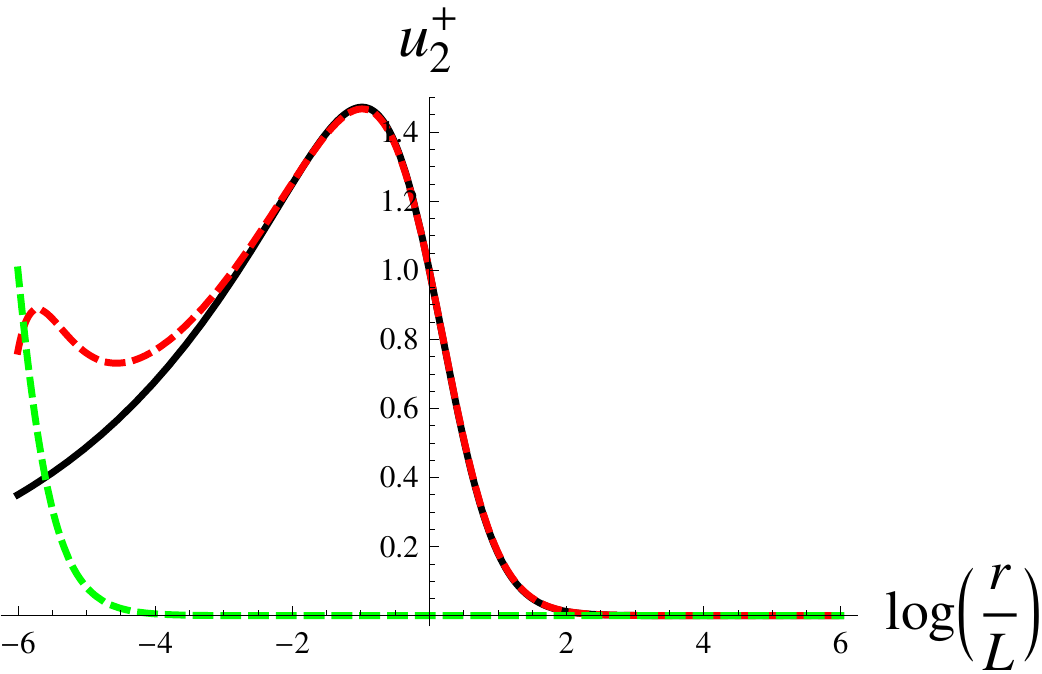}
  \caption{A normal mode with $u^+_2$ nonzero (corresponding to a pole in $G_{22}$) for $m=0$, $Q/L=1$, $q/L=2$ and $k/\Omega=3/2$. The solid line corresponds to $u^+_2$, which is purely real. For this normal mode $u^+_1$ and $u^-_2$ are exactly zero and $u^-_1$ is purely imaginary (the analytical forms are given in \eqref{GotNormalMode}). The dashed lines show the real (Red) and imaginary (Green) parts of the purely infalling solution for $L\omega=10^{-3}$. Note that they match away from $r=0$ and differ for small $r$, where the terms proportional to $\omega$ in \eqref{Diracb} become dominant.}
  \label{fig:normalmode}
\end{figure}
To find a normal mode with this choice of parameters, our approach was to numerically integrate \eqref{Diracb} with initial conditions given by the $r=0$ series expansion (with first term given by \eqref{HorSeriesZF} with $V=0$) at some small but finite $r$. The near boundary coefficients $C^+_a$ where then extracted by evaluating the Wronskian of the numerical solution with the boundary solution \eqref{uInAdS} and $k$ was varied until a zero was found.  The range allowed by \eno{kLimBdy} and \eno{kLimHor} is ${7 \over 6} \leq {k \over \Omega} \leq 2$.  
We found a normal mode with $u^+_2$ nonzero in this range and were even able to write it down in closed form! The normal mode occurs for $k/\Omega=3/2$ (see Fig.~\ref{fig:normalmode}), and is given by
\eqn{GotNormalMode}{
u^+_2 &= {2 r^{1\over 3}\over (1+r^2)^{1\over6}(2+r^2)^{5\over 4}} \left(\sqrt{2+r^2}-r\right)^{1\over 2} \cr
u^-_1 &= \sqrt{2}i {2  r^{1\over 3}\over{(1+r^2)^{1\over 6}(2+r^2)^{5\over 4}}}  \left(\sqrt{2+r^2}-r\right)^{{}-{1\over 2}} \,,
}
and $u^+_1=u^-_2=0$. It is easy to verify this is a solution of \eqref{Diracb},  and expanding it at small $r$ we obtain $u^+_2 = r^{1\over 3}+ O(r^{4\over 3})$, which shows that $V=0$ (cf. \eqref{HorSeriesZF}). If we expand it at large $r$, we find $u^+_2 = 2 r^{-3} + O(r^{-4})$, and this shows that $C^+_2=0$ (if it were nonzero, there would be a term proportional to $r^{-2}$) and so that \eqref{GotNormalMode} is in fact a normal mode.

One can understand the simple analytic form of the charged dilatonic black hole solution \eno{TwoChargeExpress} as a consequence of its relation to spinning D3-branes.  The analytic form of the normal mode \eno{GotNormalMode} is more mysterious, since neither the action \eno{FermionicS} nor the choice of parameters \eno{ParameterChoice} was drawn from string theory.

\section{A thermodynamic instability}
\label{UNSTABLE}

Now let us turn to the question of embedding \eno{Lred} and the solution \eno{TwoChargeExpress} into string theory.  We will treat this embedding in two steps.  First, in this section, we will discuss embedding of \eno{Lred} in maximal gauged supergravity.  From this embedding, one can already see the thermodynamic instability mentioned in the introduction.  Then in section~\ref{GIANT} we will consider the lift to ten dimensions.  We note in advance that only a small fraction of this section and the next is original.  The rest is a compilation of results from the literature, in particular \cite{Behrndt:1998jd,Chamblin:1999tk,Cvetic:1999rb,Cvetic:1999xp,Gubser:2004xx}.

The $SO(6)_R$ symmetry of ${\cal N}=4$ SYM corresponds to the rotation group in directions transverse to the D3-branes: that is, rotations of the $S^5$ factor in $AdS_5 \times S^5$.  The Cartan subalgebra of $SO(6)$ is $U(1)^3$.  Translationally-invariant states of ${\cal N}=4$ SYM can have three independent commuting R-charge densities corresponding to these three $U(1)$'s.  Let's denote these R-charge densities in the field theory as $\rho_i$.  As before, let $\epsilon$ and $s$ be the energy and entropy densities in the field theory, and let us define rescaled quantities
 \eqn{hattedDefs}{
  \hat\epsilon \equiv {\epsilon \over N^2} \qquad
  \hat{s} \equiv {s \over N^2} \qquad
  \hat\rho_i = {\rho_i \over N^2} \,.
 }
Because of the relation
 \eqn{Nrelation}{
  {L^3 \over \kappa^2} = {N^2 \over 4\pi^2}
 }
for $SU(N)$ ${\cal N}=4$ SYM, the definitions of $\hat\epsilon$ and $\hat{s}$ agree with the ones given in \eno{RescaledDensities}.

The $SO(6)_R$ symmetry of ${\cal N}=4$ SYM is the gauge group of maximal gauged supergravity in five dimensions.  The lagrangian of this theory is quite complicated, but a simple truncation of it that includes the $U(1)^3$ Cartan subalgebra is the STU model \cite{Behrndt:1998jd},
 \eqn{LSTU}{
  {\cal L} = {1 \over 2\kappa^2} \left[ 
    R - {1 \over 2} G_{ij} F_{\mu\nu}^i F^{\mu\nu j} - 
     G_{ij} \partial_\mu X^i \partial^\mu X^j - V(X) + 
     \hbox{Chern-Simons} \right] \,.
 }
The Chern-Simons term will not matter for the calculations of this paper.  The index $i$ runs from $1$ to $3$, and the scalars $X^i$ are constrained to satisfy $X^1 X^2 X^3 = 1$.  The target space metric and potential are given by
 \eqn{GandV}{
  G_{ij} = {1 \over 2} 
    \diag\left\{ {1 \over (X^1)^2}, {1 \over (X^2)^2}, {1 \over (X^3)^2}
    \right\} \qquad
  V = -{4 \over L^2} \sum_{i=1}^3 {1 \over X^i} \,.
 }
The general translationally invariant black brane solution with unequal electric charges is
 \eqn{GeneralThreeCharge}{
  ds^2 &= -{f \over H^{2/3}} dt^2 + {H^{1/3} \over f} dr^2 + H^{1/3}
    {r^2 \over L^2} d\vec{x}^2  \cr
  \Phi_i &= {Q_i \sqrt\mu \over r^2 + Q_i^2} - 
    {Q_i \sqrt\mu \over r_H^2 + Q_i^2} \qquad
  X^i = {H^{1/3} \over H_i}  \cr
  f &= -{\mu \over r^2} + {r^2 \over L^2} H \qquad 
   H = H_1 H_2 H_3 \qquad H_i = 1 + {Q_i^2 \over r^2}
 }
The horizon radius $r_H$ is determined as the largest root of $f$.  One straightforwardly finds
 \eqn{ThermoIdent}{
  \hat\epsilon = {3\mu \over 8\pi^2 L^6} \qquad
  \hat{s} = {r_H \sqrt\mu \over 2\pi L^5} \qquad
  \hat\rho_i = {Q_i \sqrt\mu \over 4\pi^2 L^5} \,.
 }
To recover \eno{Lred} from \eno{LSTU}, we set
 \eqn{FieldChoices}{
  A_\mu^3 = 0 \qquad A_\mu^1 = A_\mu^2 = A_\mu/\sqrt{2} \qquad
   X^1 = X^2 = e^{-2\alpha} \qquad X^3 = e^{4\alpha} \,.
 }
It is now easy to check that the solution \eno{TwoChargeExpress} and its thermodynamics \eno{RescaledDensities} are recovered from \eno{GeneralThreeCharge} and \eno{ThermoIdent} by setting
 \eqn{TwoChargeChoice}{
  Q_3 = 0 \qquad Q_1 = Q_2 = Q \qquad
  \hat\rho_3 = 0 \qquad \hat\rho_1 = \hat\rho_2 = {\hat\rho \over \sqrt{2}} \,.
 }
It may seem odd that there is no factor of $\sqrt{2}$ between $Q_1$ and $Q$.  This can be understood as a convention on the normalization of the $Q_i$.  $Q$ and the $Q_i$ are really length scales in the five-dimensional geometries, not charge densities per se, so their normalization is essentially arbitrary.

The relations \eno{ThermoIdent} parametrize the equation of state.  Following \cite{Cvetic:1999rb}, one may make this parameterization more efficient by defining
 \eqn{DefineY}{
  y = {r \over \sqrt[4]{\mu L^2}} \qquad
  y_H = {r_H \over \sqrt[4]{\mu L^2}} 
      = {3^{3/4} \over 2^{5/4} \sqrt\pi} {\hat{s} \over \hat\epsilon^{3/4}} \qquad
  y_i = {Q_i \over \sqrt[4]{\mu L^2}} 
      = {3^{3/4} \sqrt\pi \over 2^{1/4}} {\hat\rho_i \over \hat\epsilon^{3/4}} \,.
 }
Then the condition that $f=0$ at $r=r_H$ becomes
 \eqn{yHPoly}{
  (y_1^2 + y_H^2) (y_2^2 + y_H^2) (y_3^2 + y_H^2) - y_H^2 = 0 \,.
 }
The equation \eno{yHPoly} is a relation among the dimensionless ratios $\hat{s}/\hat\epsilon^{3/4}$ and $\hat\rho_i / \hat\epsilon^{3/4}$, and as such it is all the scale-invariant information available about the equation of state.  One can solve it explicitly for $y_H$, and from that solution extract an expression for $\hat{s}$ in terms of $\hat\epsilon$ and $\hat\rho_i$.  This expression is quite complicated because it involves solving a general cubic equation.  A considerably easier procedure is to define new variables
 \eqn{xiDef}{
  x_i \equiv {y_i \over y_H} = 2\pi {\hat\rho_i \over \hat{s}}
 }
and note that \eno{yHPoly} becomes
 \eqn{yHSolve}{
  {1 \over y_H^4} = (1+x_1^2)(1+x_2^2)(1+x_3^2) \,.
 }
This can be converted immediately into a simple expression for $\hat\epsilon$ in terms of $\hat{s}$ and $\hat\rho_i$.  For the two charge case, this expression is
 \eqn{TwoChargeThermo}{
  \hat\epsilon &= {3 \over 2^{5/3} \pi^{2/3}} 
    \sqrt[3]{(\hat{s}^2 + 4\pi^2 \hat\rho_1^2) (\hat{s}^2 + 4\pi^2 \hat\rho_2^2)}  \cr
    &\approx {3\pi^{2/3} \over 2^{1/3}} (\hat\rho_1 \hat\rho_2)^{2/3}
     \left[ 1 + \left( {1 \over \hat\rho_1^2} + {1 \over \hat\rho_2^2} \right)
       {\hat{s}^2 \over 12\pi^2} + O(s^4) \right] \,,
 }
where in the second line we have passed to the near-extremal limit, where the entropy density is much smaller than the charge densities.  From the fact that the leading dependence of $\hat\epsilon$ on $\hat{s}$ near extremality is quadratic, it follows immediately that the entropy density is linear in the temperature close to extremality.  More explicitly, one can solve $T = {\partial\hat\epsilon / \partial \hat{s}}$ for $\hat{s}$ to find 
 \eqn{LinearSGeneral}{
  \hat{s} \approx (2\pi)^{4/3} {(\hat\rho_1 \hat\rho_2)^{4/3} \over
    \hat\rho_1^2 + \hat\rho_2^2} T \qquad\hbox{for low $T$.}
 }
\eno{LinearSGeneral} generalizes \eno{LinearS} to the case of unequal charges.

The equation of state \eno{TwoChargeThermo} encodes a thermodynamic instability, which is a special case of the Gregory-Laflamme type instabilities found in \cite{Cvetic:1999rb}.  To see it in a simple way, let us rewrite \eno{TwoChargeThermo} as 
 \eqn{Foundehat}{
  \hat\epsilon = {3 \over 2^{5/3} \pi^{2/3}} 
    \sqrt[3]{(\hat{s}^2 + 2\pi^2 (\hat\rho-\hat\rho_z)^2) 
             (\hat{s}^2 + 2\pi^2 (\hat\rho+\hat\rho_z)^2)} \,,
 }
where we have defined
 \eqn{rhoDefs}{
  \begin{pmatrix} \hat\rho \\ \hat\rho_z \end{pmatrix} = 
    {1 \over \sqrt{2}} \begin{pmatrix} 1 & 1 \\ -1 & 1 \end{pmatrix} 
    \begin{pmatrix} \hat\rho_1 \\ \hat\rho_2 \end{pmatrix} \,.
 }

Local thermodynamic stability is the condition that $\hat\epsilon$ is concave up as a function of the extensive thermodynamic variables $\hat{s}$ and $\hat\rho_i$.  One therefore calculates the Hessian matrix
 \eqn{ExplicitHessian}{
  {\partial^2 \hat\epsilon \over \partial(\hat{s},\hat\rho,\hat\rho_z)^2} \equiv
    \begin{pmatrix} {\partial^2 \hat\epsilon \over \partial\hat{s}^2} & 
      {\partial^2 \hat\epsilon \over \partial\hat{s} \partial\hat\rho} &
      {\partial^2 \hat\epsilon \over \partial\hat{s} \partial\hat\rho_z}  \\
     {\partial^2 \hat\epsilon \over \partial\hat\rho \partial\hat{s}} & 
      {\partial^2 \hat\epsilon \over \partial\hat\rho^2} &
      {\partial^2 \hat\epsilon \over \partial\hat\rho \partial\hat\rho_z}  \\
      {\partial^2 \hat\epsilon \over \partial\hat\rho_z \partial\hat{s}} & 
     {\partial^2 \hat\epsilon \over \partial\hat\rho_z \partial\hat\rho} &
      {\partial^2 \hat\epsilon \over \partial\hat\rho_z^2}
    \end{pmatrix} = 
   {3 \over 8} \left( {\hat{s} \over \hat\epsilon} \right)^2
    \begin{pmatrix} {2+3x_\rho^2 \over 2\pi^2} & -{2x_\rho \over \pi} & 0  \\
     -{2x_\rho \over \pi} & 6 + x_\rho^2 & 0  \\
     0 & 0 & 6 - 3 x_\rho^2 \end{pmatrix} \,,
 }
where the last expression is valid only for $\hat\rho_z = 0$, and we have defined
 \eqn{xRhoDef}{
  x_\rho \equiv 2\pi {\hat\rho \over \hat{s}} \,.
 }
The $(\hat{s},\hat\rho)$ block of the Hessian matrix is positive definite for all values of $x_\rho$, but evidently $\partial^2 \hat\epsilon / \partial\hat\rho_z^2$ goes smoothly from positive to negative values as $x_\rho$ increases through the value $\sqrt{2}$.  Thus $x_\rho = \sqrt{2}$ is the boundary of thermodynamic stability, and the instability arises on the low temperature, high charge density side of it.  Correspondingly, if one defines conjugate variables $\Omega$ and $\Omega_z$ dual to $\hat\rho$ and $\hat\rho_z$, and also a rescaled free energy density
 \eqn{fDef}{
  \hat{f} = \hat\epsilon - T\hat{s} - \Omega\hat\rho - \Omega_z \hat\rho_z \,,
 }
then the matrix of susceptibilities $\partial^2 \hat{f} / \partial(T,\Omega,\Omega_z)^2$ is minus the inverse of $\partial^2 \hat\epsilon / \partial(\hat{s},\hat\rho,\hat\rho_z)^2$.  In particular, the $\hat\rho_z$ susceptibility is
 \eqn{ChiZDef}{
  \hat\chi_z \equiv -{\partial^2 \hat{f} \over \partial\Omega_z^2} = 
    {1 \over \partial^2 \hat\epsilon / \partial\hat\rho_z^2} = 
     {8 \over 9} \left( {\hat\epsilon \over \hat{s}} \right)^2 
      {1 \over 2-x_\rho^2} = {4 \over 9} {\hat\epsilon^2 \over \hat{s}^2 - 
        2\pi^2 \hat\rho^2} \approx {\sqrt{2} \, \hat\rho / 3\pi \over T-T_c} \,,
 }
where 
 \eqn{TcValue}{
  T_c = {\Omega \over \sqrt{2}\, \pi} \,.
 }
The approximate equality in \eno{ChiZDef} is accurate near $T_c$.  The behavior of $\hat\chi_z$ is reminiscent of mean-field ferromagnetism.

It is worth remarking that the off-diagonal charge $\hat\rho_z$ is actually one component of an $SU(2)$ triplet of charges, call them $(\hat\rho_x,\hat\rho_y,\hat\rho_z)$.  To see this, recall first that that $U(1)^3$ charges $(\hat\rho_1,\hat\rho_2,\hat\rho_3)$ are embedded in $SU(4)_R \approx SO(6)_R$ as the Cartan subalgebra.  When all of them are non-zero and equal (i.e.~for RN$AdS_5$), a symmetry $U(3) \subset SU(4)_R$ is preserved.  This $U(3)$ is the one which acts on the three complex scalars of ${\cal N}=4$ SYM as a triplet.  When one sets $\hat\rho_3=0$, this $U(3)$ is broken to $U(2) \times U(1)$.  The $U(1)$ inside $U(2)$ (that is, its center) corresponds to the charge $\hat\rho$, while the $J_z$ generator of $SU(2)$ (in a conventional Pauli basis) corresponds to $\hat\rho_z$.  The other generators of $SU(2)$ correspond to $\hat\rho_x$ and $\hat\rho_y$.  Note that the $SU(2)$ under discussion is a global flavor symmetry in the boundary theory, making it more similar to isospin than to intrinsic angular momentum of particles.  The susceptibilities $\hat\chi_x$ and $\hat\chi_y$ are identical to $\hat\chi_z$ when $\hat\rho_x=\hat\rho_y=\hat\rho_z=0$.  So the instability \eno{ChiZDef} involves a spontaneous breaking of the $SU(2)$ symmetry.  One should expect the typical state to involve non-zero $(\hat\rho_x,\hat\rho_y,\hat\rho_z)$ pointing in different directions in different parts of space.  Topological defects may even be possible.  The main difficulty is that there is little to no understanding about the endpoint of evolution of the thermodynamic instability.  At linear order, in a region where only $\hat\rho_z$ becomes non-zero, what is happening is that $\hat\rho_1$ and $\hat\rho_2$ become slightly unequal without otherwise changing the thermodynamics.

As explained in \cite{Gubser:2000ec,Gubser:2000mm}, thermodynamic instabilities correspond to dynamical instabilities in the real-time geometry.  In the present case, where the instability relates (at linear order) only to the development of an $SU(2)$ charge and not to a spatial modulation of the entropy density, the dynamical instability should involve only matter fields in the bulk, not the metric.  The pertinent matter fields should be the $SU(2)$ gauge field and certain scalars that couple to it through its kinetic term.  In the case of the instability toward developing non-zero $\hat\rho_z$, the only gauge field involved is $A_\mu^{1}-A_\mu^{2}$, and the only scalar involved is $X^1/X^2$.  Because the instability relates to thermodynamics, it should be present at arbitrarily small wave-number.  Presumably it ceases to exist at a critical wave-number, $k_{\rm GL}$, comparable to the chemical potential $\Omega$.

\section{A ten-dimensional lift}
\label{GIANT}

As we have already remarked, the charged dilatonic black hole has a naked singularity at $r=0$.  We should study the ten-dimensional lift via maximal gauged supergravity to find out if this singularity has an obvious resolution.  This lift is well known \cite{Cvetic:1999xp}: for the general three-charge black hole \eno{GeneralThreeCharge}, it is
 \eqn{TenLift}{
  ds_{10}^2 &= \sqrt\Delta \left[ -{f \over H} dt^2 + {dr^2 \over f} + 
    {r^2 \over L^2} d\vec{x}^2 \right] + 
   {1 \over \sqrt\Delta} \sum_{i=1}^3 H_i \left(
     L^2 d\mu_i^2 + \mu_i^2 \left[ L d\phi_i + {\sqrt\mu \over Q_i} 
       \left( {1 \over H_i} - 1 \right) dt \right]^2 \right)  \cr
  F_5 &= G_5 + * G_5 \qquad G_5 = dB_4 \qquad
   B_4 = -{r^4 \over L^4} \Delta dt \wedge d^3 x - 
     \sum_{i=1}^3 {Q_i \sqrt\mu \over L} \mu_i^2 d\phi_i \wedge d^3 x \,.
 }
In addition to the functions $f$, $H$, and $H_i$ appearing in \eno{GeneralThreeCharge}, we have defined
 \eqn{SomeDefs}{
  \Delta = H \sum_{i=1}^3 {\mu_i^2 \over H_i} \qquad
  \mu_1 = \cos\theta_1 \cos\theta_2 \qquad
  \mu_2 = \cos\theta_1 \sin\theta_2 \qquad
  \mu_3 = \sin\theta_1 \,.
 }
Evidently, $\mu_1^2 + \mu_2^2 + \mu_3^2 = 1$, so $(\theta_1,\theta_2)$ are coordinates on $S^2$.  The $S^5$, parametrized by $(\theta_1,\theta_2,\phi_1,\phi_2,\phi_3)$, is thus regarded as a $T^3$ fibration over $S^2$.  If we set $Q_1=Q_2=Q$, $Q_3=0$, and $\mu = Q^4/L^2$ (the last corresponding to the extremal limit) and approach the horizon at $r=0$, the solution takes the following form:
 \eqn{TenNear}{
  ds_{10,\rm near}^2 &= |\mu_3| \left( -{2r^2 \over L^2} dt^2 + 
      {L^2 \over 2r^2} dr^2 + {r^2 L^2 \over Q^2} d\phi_3^2 \right) + 
    |\mu_3| {Q^2 \over L^2} d\vec{x}^2 
    \cr &\qquad{} 
    + {1 \over |\mu_3|} \left[ L^2 d\mu_1^2 + L^2 d\mu_2^2 + 
      \mu_1^2 \left( Ld\phi_1 - {Q \over L} dt \right)^2 + 
      \mu_2^2 \left( Ld\phi_2 - {Q \over L} dt \right)^2 \right]  \cr
  B_{4,\rm near} &= -{Q^3 \over L^3} 
   \left[ \mu_1^2 \left( Ld\phi_1 - {Q \over L} dt \right) + 
          \mu_2^2 \left( Ld\phi_2 - {Q \over L} dt \right) \right] \wedge d^3 x \,.
 }
Intriguingly, the metric has an $AdS_3$ factor, suggesting that the infrared dynamics is controlled by a 1+1-dimensional conformal field theory.  The rest of the metric, on any given time slice, is a sum of two conformally flat pieces, where the conformal factors, $|\mu_3| = \sqrt{1 - \mu_1^2 - \mu_2^2}$ or its reciprocal, are finite and non-zero only on the interior of the unit disk in the $\mu_1$-$\mu_2$ plane.  The locus of points where $\mu_3=0$, and where the metric \eno{TenNear} is ill-defined, corresponds to the equator of $S^2$, and to an equatorial $S^3$ of $S^5$.  The geometry \eno{TenNear} in fact splits into two identical pieces, joined along this $S^3$.  Each is a warped product of $AdS_3$, ${\bf R}^3$, and the unit ball in ${\bf R}^4$, where the unit ball spins rigidly and equally in the two independent planes of ${\bf R}^4$.

It is not entirely surprising to find an $AdS_3$ factor in the near-horizon geometry.  Following \cite{McGreevy:2000cw}, we note that angular momentum in the $S^5$ directions can be carried by D3-branes wrapped on various three-spheres in that $S^5$.  These wrapped D3-branes, called ``giant gravitons,'' have as their field theory dual operators which are determinants or sub-determinants of one of the three complex adjoint scalar fields of ${\cal N}=4$ SYM \cite{Balasubramanian:2001nh}, which we will denote $Z_1$, $Z_2$, and $Z_3$.  More specifically: $\det Z_i$ corresponds to a D3-brane wrapped on the equatorial $S^3$ of $S^5$ found by intersecting the hyperplane $z_i=0$ with the sphere $\sum_{j=1}^3 |z_j|^2 = L^2$ in ${\bf C}^3$.  It was suggested in \cite{Balasubramanian:2001dx} that a large enough number of giant gravitons of one type (say, the type associated with $Z_1$) could be described by an $AdS_5 \times S^5$ geometry that they create around themselves.  Now consider a configuration with both $\det Z_1$ and $\det Z_2$ giant gravitons.  They intersect along an equatorial $S^1$ of $S^5$, defined by intersecting $z_1=z_2=0$ with $\sum_{i=1}^3 |z_i|^2 = L^2$.  The low-energy dynamics at the intersection is plausibly a 1+1-dimensional CFT whose central charge is proportional to the product of the number of giant gravitons of each type, in analogy to how D3-branes on orthogonal cycles of $T^4$ intersect over the ``effective string'' of \cite{Strominger:1996sh,Callan:1996dv}.  This CFT, we presume, is dual to the $AdS_3$ factor in \eno{TenNear}.

The full story is presumably somewhat more complicated than the intersecting equatorial D3-branes described in the previous paragraph.  In particular, it is known \cite{Myers:2001aq} that a singular, supersymmetric relative of the solutions \eno{TenLift} can be understood in terms of three orthogonal sets of giant gravitons with a distribution of sizes.  Moreover, in the treatment of \cite{Myers:2001aq}, black holes with horizons have finite $\mu$, and so are finitely far from the supersymmetric case, where $\mu=0$.

The presence of an $AdS_3$ factor fits nicely with the linear specific heat, since all $1+1$-dimensional CFT's have linear specific heat: that's simply on account of the fact that entropy density, as a dimension one object, must scale linearly with the only available energy scale, namely temperature.  On the gravity side, the near-extremal generalization of \eno{TenNear} involves the $AdS_3$-Schwarzschild geometry:
 \eqn{NEGTen}{
  ds_{10,\rm near}^2 &= |\mu_3| \left( -{2(r^2-r_H^2) \over L^2} dt^2 + 
      {L^2 \over 2(r^2-r_H^2)} dr^2 + {r^2 L^2 \over Q^2} d\phi_3^2 \right) + 
    |\mu_3| {Q^2 \over L^2} d\vec{x}^2 
    \cr &\qquad{} 
    + {1 \over |\mu_3|} \left[ L^2 d\mu_1^2 + L^2 d\mu_2^2 + 
      \mu_1^2 \left( Ld\phi_1 - {Q \over L} dt \right)^2 + 
      \mu_2^2 \left( Ld\phi_2 - {Q \over L} dt \right)^2 \right] \,,
 }
with $B_4$ unchanged from \eno{TenNear}.  Like \eno{TenNear}, \eno{NEGTen} is not only a limiting form of a solution of the equations of motion of type~IIB supergravity; it is by itself a solution of those equations, away from $\mu_3=0$.

From the $1+1$-dimensional CFT perspective, the ground state entropy of the three-charge black hole in $AdS_5$ arises from partitioning a specified amount of momentum along the effective string into excitations of one chiral half of the theory, while leaving the other chiral half in its vacuum state.  The Virasoro algebra acting on the $AdS_2$ factor of the near-horizon RN$AdS_5$ geometry is probably the same as the one from the chiral half of the effective string CFT that remained in its ground state.

In our discussion of ten-dimensional lifts, we seem to have committed ourselves to the maximally supersymmetric case, which seems unlikely to have much to do with Fermi liquids given that there are massless charged scalars in the gauge theory.  Other lifts to ten dimensions may be possible, with reduced supersymmetry, as in \cite{Gubser:2009qm}; and some of these lifts could be dual to a CFT with no charged scalars, or no gauge-invariant operators built from charged scalars that can condense.  However, a fermionic field theory description may be closer to hand than it appears in the maximally supersymmetric case: at least in the case of a single charge, one can show \cite{Lin:2004nb} that BPS solutions are classified in terms of free fermions.  Our limiting solution \eno{TenNear} is fairly similar to the ansatz of \cite{Lin:2004nb}. It is tempting to think it preserves some fraction of supersymmetry, and that there might be an analogous fermionic description.  If there is such a description, the fermions probably emerge from the dynamics of eigenvalues of large $N$ matrices, as in the $c=1$ matrix model.

\section{An $AdS_4$ example}
\label{FOUR}

An obvious extension of the ideas discussed in the previous section to the case of $AdS_4 \times S^7$ is to consider three mutually orthogonal groups of M5-brane giant gravitons, each on an equatorial $S^5$ in $S^7$.  The triple intersection of M5-branes from each group is an equatorial $S^1$ in $S^7$.  This intersection is similar to the effective string of \cite{Maldacena:1997de}, and plausibly the low-energy dynamics is a 1+1-dimensional CFT.  So one expects in the eleven-dimensional lift of the extremal three-charge $AdS_4$ black hole to find an $AdS_3$ factor similar to the form \eno{TenNear}.  One also expects linear specific heat at low temperature.  This last expectation is easy to check using just the thermodynamic formulas of \cite{Cvetic:1999rb}---and it is true. 
Let us consider the general eleven-dimensional geometry, given by \cite{Cvetic:1999xp}
 \eqn{ElevenLift}{
  ds_{11}^2 &= \tilde\Delta^{2\over 3} \left[ -{\tilde f \over \sqrt{\tilde H}} dt^2 + {\tilde r^2\over 4 L^2}{\sqrt{\tilde H}\over f} d \tilde r^2  + 
   {r^4 \over   4 L^4} \sqrt{\tilde H} d\vec{x}^2 \right]  \cr &\qquad+
   {4\over {\tilde\Delta}^{{1\over 3}}\tilde H^{1\over 4}} \sum_{i=1}^4 \tilde{H}_i \left(
     L^2 d\mu_i^2 + \mu_i^2 \left[ L d\phi_i - {\sqrt {\mu \over 4 Q_i} } 
       \left( {1 \over H_i} - 1 \right) dt \right]^2 \right)  \cr
  F_4 &= d A_3  \qquad A_3 = \ - {r^6 \over (2L)^6} {\tilde H}^{3\over 4} {\tilde\Delta} dt \wedge d^2 x
  + \sum_{i=1}^4 {2\sqrt {\mu Q_i}}\, \mu_i^2 d\phi_i \wedge d^2 x \,.
}
Here, the $\mu_i$ parametrize an $S^3$, i.e., $\mu_1^2+\mu_2^2+\mu_3^2+\mu_4^2=1$, while the $\phi_i$ parametrize a $T^4$. Together, these coordinates cover an $S^7$, regarded as a $T^4$ fibration over $S^3$.  We introduced the functions
\eqn{ElevenLiftFunctions}{
\tilde H_i = 1+ {4 L Q_i\over \tilde r^2} \qquad \tilde H = \tilde H_1 \tilde H_2 \tilde H_3 \tilde H_4 \qquad \tilde\Delta = \tilde H^{1/4} \sum_{i=1}^4 {1\over \tilde H_i} \qquad
\tilde f = - {\tilde\mu\over \tilde r} +  {\tilde r^2\over L^2} \tilde H \,.
}
Note that \eqref{ElevenLift} approaches $AdS_4\times S^7$ for large $\tilde r$, and this is most easily seen by using the radial coordinate $r = {\tilde r^2\over 2 L}$. We are most interested in the three-equal-charge case $Q_1=Q_2=Q_3=Q$ and $Q=0$. The extremal limit then corresponds to taking $\tilde\mu={Q^6\over 2 L^5}$, and near horizon limit of the resulting geometry is given by
\eqn{ElevenNear}{
  ds_{11,\rm near}^2 &= |\mu_4|^{4/3} \left( -{3 Q \tilde r^2\over 4 L^3 } dt^2 + 
      {4 L^2\over 3 r^2} dr^2 + {L r^2\over Q} d\phi_4^2 \right)   + 
    |\mu_4|^{4/3} {Q^2\over  L^2} d\vec{x}^2 \cr &\quad 
    + {4 \over  |\mu_4|^{2/3}} \Bigg[ \sum_{i=1}^3 L^2 d\mu_i^2 +  \mu_i^2 \left( L d\phi_i + {Q \over 2 L} dt \right)^2\Bigg]  \cr
      A_{3,\rm near} &= {Q^3 \over L^3s} \sum_{i=1}^3 \mu_i^2 \left[  dt + {2 L^2\over Q} d\phi_i \right]\wedge d^2 x \,,
}
where we emphasize that the sums extend to $i=3$ only. As expected, we find an $AdS_3$ factor.

When reduced to four dimensions, the geometry \eqref{ElevenLift} gives rise to multiply charged black holes. The minimal four dimensional lagrangian that has the three-equal-charge black hole we are interested in as a solution is \cite{Gubser:2000mm}
\eqn{L4dim}{
  {\cal L} = {1 \over 2\kappa^2} \left[ R - {1 \over 4} e^{\alpha} F_{\mu\nu}^2 - 
    {3\over 2} (\partial_\mu\alpha)^2 + {6 \over L^2} \cosh \alpha \right]\,,
    }
and the four dimensional geometry  is given by
\eqn{ThreeChargeM2}{
  ds^2 &= e^{2A} (-h dt^2 + d\vec{x}^2) + {e^{2B} \over h} dr^2 \qquad F= dA\qquad
    A_\mu dx^\mu = \Phi dt  \cr
  A &= \log {r \over L} + {3 \over 4} \log\left( 1 + {Q \over r} \right) \qquad
  B = - A  \qquad  h = 1 - {\mu L^2\over (Q+r)^3} \cr
   \alpha &= {1\over 2} \log\left(1+ {Q\over r}\right) \qquad
  \Phi = {\sqrt{3 Q \mu}  \over Q+r} - {\sqrt{3 Q} \mu^{1\over 6} \over L^{2\over 3}} \,. 
  }
The extremal limit of \eqref{ThreeChargeM2} now corresponds to taking $\mu=Q^3/L^2$.  If we consider a four dimensional Dirac fermion in the extremal limit of\eqref{ThreeChargeM2}, and with action given by the four dimensional analogue of \eqref{FermionicS}, we again find an isolated normal mode. With the conventions of \cite{Iqbal:2009fd} for the $\Gamma$ matrices and the ansatz \eqref{PsiAnsatz} for $\psi$, Dirac's equation can be written
\eqn{Dirac4d}{
\left[ e^{-B}\sqrt{h}\Gamma^{\underline{r}} \partial_r +  ie^{-A}\left(k \Gamma^{\underline{1}} - \frac{\omega +q \Phi}{\sqrt h} \Gamma^{\underline{0}}\right) +\frac{6 h A' +h'}{4\sqrt h e^B}\Gamma^{\underline{r}} 
- m\right] u =0 \,.
}
It is useful to define the chemical potential $\tilde\Omega\equiv \sqrt{3} Q/L^2= - \Phi(+\infty)/L$. For $\omega=0$, the solutions of \eqref{Dirac4d} are approximated near $r=0$ by
\eqn{HorSeriesZF4d}{
u = U r^{-{5\over 8} + {|k| \over|\tilde\Omega|}}\bigg(1+O\big(r^{1/4}\big)\bigg)+V r^{-{5\over 8} - {|k|\over |\tilde\Omega|}}\bigg(1+O\big(r^{1/4}\big)\bigg) \qquad (\omega=0)\,,
}
and as before, we impose the boundary condition $V=0$. Near the boundary, we now have
\eqn{uInAdS4}{
u^{\pm}_a = C^{\pm}_a e^{-2 A}  X_{\mp{1\over 2} -m L}\Big(\kappa e^{-A}\Big) + D^{\pm}_a e^{-2 A}  X_{\pm{1\over 2} +m L}\Big(\kappa e^{-A}\Big) \,,
}
with $\kappa$ and $X_\nu$ the same as in the five dimensional case. Normal modes are then solutions of \eqref{Dirac4d} with $V=0$ and $C^+_a=0$, and correspond to poles of the Green's function. They can easily be found numerically, for example, in units where $L=1$, we take $m=0$, $Q=1$ (corresponding to $\tilde\Omega=\sqrt{3}$) and $q=2 $ and find a normal mode with $u^+_2$ nonzero for $k/\tilde\Omega\approx 1.26746$.

\section{Scaling solutions}
\label{SCALING}

Truncations of supergravity actions to abelian gauge fields plus neutral scalars often take the form
 \eqn{Lcommon}{
  {\cal L} = {1 \over 2\kappa^2} \left[ R - 
    {1 \over 4} \sum_a f_a(\vec\phi) (F_{\mu\nu}^a)^2 - 
    {1 \over 2} \partial_\mu \vec\phi \cdot \partial^\mu \vec\phi - 
    V(\vec\phi) \right] \,,
 }
where the functions $f_a(\vec\phi)$ and $V(\vec\phi)$ are linear combinations of exponentials of the form $e^{\vec\beta \cdot \vec\phi}$, where $\vec\beta$ is a vector of constants.  Examples involving only scalars were explored in some depth in \cite{Gubser:2000nd}.  There it was argued that in typical near-extremal solutions, the scalars run away in a definite direction where one term in $V(\vec\phi)$ dominates, or where several terms dominate over all others and stand in fixed ratio with themselves.  In other words, to understand typical near-extremal dynamics, it is enough to consider a single canonically normalized real scalar, call it $\phi$, with a potential
 \eqn{VChoice}{
  V(\phi) = V_0 e^{\eta\phi} \,,
 }
where $\eta$ is a positive constant and $V_0$ is negative, and $\phi$ is assumed to diverge to $+\infty$ in the extremal solution.  It was argued there that the entropy density scales at low temperature as $T^\chi$, where $\chi = 24/(8-3\eta^2)$ when the bulk geometry is five-dimensional.  In this section, we will study charged black holes solutions to the lagrangian
 \eqn{SingleL}{
  {\cal L} = {1 \over 2\kappa^2} \left[ R - {f(\phi) \over 4} F_{\mu\nu}^2 - 
    {1 \over 2} (\partial\phi)^2 - V(\phi) \right] \,,
 }
where $V(\phi)$ is chosen as in \eno{VChoice}, and
 \eqn{fChoice}{
  f(\phi) = e^{\gamma\phi}
 }
for some constant $\gamma$.  We again assume that the bulk is five-dimensional.  The large $\alpha$ behavior of the theory \eno{Lred} that we started with corresponds to $\eta = 1/\sqrt{6}$ and $\gamma = \sqrt{2/3}$.  Because $\eta$ and $\gamma$ are defined in reference to a canonically normalized scalar, their values would change if the scalar kinetic term changes.  Renormalization of scalar kinetic terms is commonplace in theories without a high degree of supersymmetry.  So it makes sense to work out what happens for arbitrary $\eta$ and $\gamma$.

We start with an ansatz closely related to \eno{TwoChargeExpress}:
 \eqn{MetricAnsatz}{
  ds^2 = L^2 \left[ e^{2A} \left( -h \, dt^2 + d\vec{x}^2 \right) + {dr^2 \over h}
    \right] \qquad
  A_\mu dx^\mu = L \Phi \, dt \,,
 }
where it is assumed that $A$, $h$, $\Phi$, and $\phi$ depend only on $r$.  $L$ is, at this stage, an arbitrary length scale, present in order to render the coordinates $(t,\vec{x},r)$ dimensionless.  Following \cite{Gubser:2009cg}, one can construct a Noether charge
 \eqn{NoetherCharge}{
  {\cal Q} \equiv e^{2A} \left( e^{2A} h' - f(\phi) \Phi \Phi' \right) \,.
 }
${\cal Q}$ is independent of $r$ when the equations of motion are obeyed.  Moreover, evaluating ${\cal Q}$ at the horizon, where $\Phi$ is required to vanish, shows that ${\cal Q} = 2\kappa^2Ts$. So ${\cal Q}=0$ is an extremality condition.

The Maxwell equation can be solved directly, leading to
 \eqn{FoundPhip}{
  \Phi' = {\rho \over e^{2A} f(\phi)} \,,
 }
where $\rho$ is a dimensionless version of charge density.  The result \eno{FoundPhip} holds for any choice of $f(\phi)$.  When we make the simple choices \eno{VChoice} and \eno{fChoice}, the following solutions to the equations of motion can be found by inspection:
 \eqn{ScalingSolution}{
  A &= {1 \over 6} (\gamma+\eta)^2 \log r  \cr
  \phi &= -(\gamma+\eta) \log r  \cr
  \Phi &= {\rho \over \zeta} (r^\zeta - r_H^\zeta)  \cr
  h &= -{2L^2 V_0 \over \zeta (2+\gamma^2+\gamma\eta)}
    r^{1-2(\gamma+\eta)^2/3} (r^\zeta - r_H^\zeta) \,,
 }
where we impose the relations
 \eqn{ZetaAndRho}{
  \zeta = 1 + {2\gamma^2+\gamma\eta-\eta^2 \over 3} \qquad
  \rho = \sqrt{-2 V_0 L^2 {2-\gamma\eta-\eta^2 \over 2+\gamma^2+\gamma\eta}} \,.
 }
It is easy to check that the solution \eno{ScalingSolution} is extremal, in the sense of having ${\cal Q}=0$, precisely when $r_H=0$.  When it is non-extremal, one can obtain the following expressions for the horizon entropy density (measured with respect to the dimensionless coordinates $\vec{x}$) and the temperature (measured with respect to the dimensionless time $t$):
 \eqn{sTvalues}{
  s = {2\pi L^3 \over \kappa^2} r_H^{(\gamma+\eta)^2/2} \qquad
  T = -{L^2 V_0 \over 2\pi} {r_H^{1+(\gamma-5\eta)(\gamma+\eta)/6} \over
    2 + \gamma^2 + \gamma\eta} \,.
 }
Evidently, we have the scaling behavior
 \eqn{sScales}{
  s \propto T^\chi \qquad\hbox{where}\qquad
   \chi = {(\gamma+\eta)^2 \over 2 + (\gamma-5\eta)(\gamma+\eta)/3} \,.
 }
It is easy to check that plugging in $\eta = 1/\sqrt{6}$ and $\gamma = \sqrt{2/3}$ gives $\chi=1$.  It should be possible to get a wide range of dependences of $s$ on $T$ by choosing more general $V(\phi)$ and $f(\phi)$.  For example, one can probably get log-corrected power-law scaling of $s$ on $T$ by altering $V(\phi)$ and/or $f(\phi)$ by powers of $\phi$.

The solutions \eno{ScalingSolution} are not asymptotically $AdS_5$.  In fact, their asymptotics for large $r$ can be rather peculiar, with $h$ growing faster than $e^{2A}$.  But if a term is added to $V(\phi)$ which makes $\phi=0$ its global maximum, we expect that asymptotically $AdS_5$ charged dilatonic black hole solutions will exist whose entropy scales as we have found in \eno{sScales} near extremality.  Probably a wide range of such black holes support fermion normal modes similar to the one described in sections~\ref{FERMION} and~\ref{FOUR}, and so are candidates for holographic duals of generalized Fermi liquids.

\section{Summary}
\label{SUMMARY}

Let's summarize the properties of the charged dilatonic black hole in $AdS_5$, given explicitly by \eno{TwoChargeExpress}:
 \begin{itemize}
  \item It has linear specific heat at low temperature, like a Fermi gas does.
  \item It supports a fermion normal mode of the type previously argued to be associated with a Fermi surface.  For the choice of parameters we made at the end of section~\ref{FERMION}, the wave-function for the normal mode has a simple closed form.
  \item Its embedding in $d=5$ maximal gauged supergravity has a thermodynamic instability toward the development of an $SU(2)$ charge, with a susceptibility that has a singularity like the Curie-Weiss law at finite temperature.  This instability is of the Gregory-Laflamme type, and the endpoint of its evolution is unknown.
  \item The five-dimensional solution has a naked singularity.
  \item The lift to ten dimensions based on $d=5$ maximal supergravity describes spinning D3-branes with two of the independent spins equal and the third zero.
  \item The ten-dimensional geometry near extremality has an $AdS_3$-Schwarzschild factor which accounts for the linear specific heat and suggests invariance of the infrared dynamics under a non-chiral Virasoro algebra.
  \item There is an obvious generalization to $AdS_4$ which realizes most of the same properties.
  \item An analysis of scaling solutions indicates that power-law behaviors $s \propto T^\chi$ with continuously variable $\chi$ can be arranged through simple modifications of the theory on which our original example was based.
 \end{itemize}
The solution \eno{TwoChargeExpress} is non-supersymmetric, except when $\mu=0$, which is finitely below the extremal limit.  The simple form of the near-horizon limit of the $10$-dimensional lift suggests that some supersymmetry (perhaps a quarter) may be recovered in this limit.  If so, the infrared dynamics would be controlled by a $1+1$-dimensional superconformal field theory.  We hope that a better understanding of the dual field theory in the asymptotically $AdS_5 \times S^5$ case can be achieved by studying states with two large commuting $R$-charges.  In a more general setting, the hope is that an explicit construction could be found where the $U(1)$ charge of the black hole is carried only by fermions in the field theory dual.  It would be particularly interesting if the exact expression \eno{GotNormalMode} we found for the normal mode generalizes to an exact two-point function, because then one would have some analytic control over excitations of a strongly interacting degenerate Fermi liquid in four spacetime dimensions.

\section*{Acknowledgments}

We thank D.~Huse and S.~Sondhi for useful discussions. This work was supported in part in part by the DOE under Grant No.\ DE-FG02-91ER40671 and by the NSF under award number PHY-0652782. FDR was also supported in part by the FCT grant SFRH/BD/30374/2006. 

\bibliographystyle{ssg}
\bibliography{fermion}

\end{document}